\documentclass{article}

\PassOptionsToPackage{numbers, compress}{natbib}
\usepackage[preprint]{neurips_2026}

\usepackage[utf8]{inputenc} 
\usepackage[T1]{fontenc}    
\usepackage{hyperref}       
\usepackage{url}            
\usepackage{booktabs}       
\usepackage{amsfonts}       
\usepackage{nicefrac}       
\usepackage{microtype}      
\usepackage{xcolor}         
\usepackage{graphicx}
\usepackage{amsmath}

\usepackage{amsthm}
\newtheorem{proposition}{Proposition}

\title{Social Theory Should Be a Structural Prior for Agentic AI: A Formal Framework for Multi-Agent Social Systems}

\author{
    \textbf{Lynnette Hui Xian Ng}$^1$ \quad 
    \textbf{Iain J. Cruickshank}$^1$ \quad
    \textbf{Adrian Xuan Wei Lim}$^2$ \quad
    \textbf{Kathleen M. Carley}$^1$
    \vspace{0.4em} \\
    \textnormal{$^1$Carnegie Mellon University} \textnormal{ $^2$National University of Singapore} \\
    \texttt{lynnetteng@cmu.edu}
}

\begin{document}

\maketitle

\begin{abstract}
Agentic AI systems are increasingly deployed not in isolation, but inside social environments populated by other agents and humans, such as in social media platforms, multi-agent LLM pipelines or autonomous robotics fleets. In these settings, system behavior emerges not from individual agents alone, but from the multi-agent interactions over time. Emergent dynamics of individuals in a social group have been long studied by social scientists in human contexts. \textbf{This position paper argues that agentic AI systems must be modeled with social theory as a structural prior, and formalizes a Multi-Agent Social Systems (MASS) framework for how agents interact and influence to generate system-level outcomes.} We represent MASS as a class of dynamical system of information generation, local influence and interaction structure, formulated by four structural priors anchored in social theory: strategic heterogeneity, networked-constrained dependence, co-evolution and distributional instability. We demonstrate the importance of each structural prior through formal propositions, and articulate a research agenda for how MASS should be modeled, evaluated and governed.
\end{abstract}

\section{Introduction}
Agentic AI systems are increasingly embedded in social environments, where their behavior is shaped not only by their individual capabilities but by interactions with other agents - and humans - over time. In these settings, outcomes are not determined only by the performance of a single agent, but emerge from the patterns of interactions across agent-agent and agent-human populations within structured networks. We have built agents capable of reasoning, planning and negotiation \cite{masterman2024landscape,vaccaro2025advancing,li2025measuring}, but have largely studied them in terms of individual task completion or as independent units contributing to a collective objective. A recent NeurIPS position paper argues ``Large Language Models Miss the Multi-Agent Mark" because of the failure to account for population dynamics \cite{la2025large}; indeed, today's agents have mostly been single-task agentic systems, and emerging social groups remain unstudied. 

The social environments that today's AI agents are deployed in include social media platforms, multi-agent LLM pipelines, autonomous robot fleets, and so forth. 
As such systems scale and become more interconnected, understanding their behavior moves beyond isolated, task-centric perspective and towards the study of interacting agents and emergent social dynamics.
The study of such social dynamics is not a new problem for science. It is what social theory has studied for decades in human populations: how opinions diffuse \cite{assenova2018modeling,zhu2018effect}, how network position governs influence \cite{granovetter1973strength,marsden1993network}, how the information environment itself is socially produced \cite{mccombs1972agenda,coleman2009agenda}. What is new is that in today's world, machines are participants, not just observers, in these environments. Given these new realities, the agentic AI field should leverage social theory to design these environments. At the same time, social theory must evolve from a science of human-only populations to a science of composite populations, where humans and machine agents interact, shape one another, and together form a social world. 

\textbf{Position.} We argue that agentic AI systems should be designed with social theory as a \emph{structural prior}. Social theory's core constructs like role differentiation \cite{merton1957role}, the co-evolution of agents \cite{centola2010spread,friedkin1997social} specify the space of interactions, observations and system trajectories. These constructs also define agents perceptions, actions, and collective behavior, and their assumptions should be built into agentic AI system architectures. We formalize this perspective through \textbf{Multi-Agent Social Systems (MASS)}, a class of dynamic networked systems in which agents (human and artificial) interact through structured, social relationships. In MASS, system-level outcomes arise from the joint dynamics of agent states, message exchange, and network structure. 



\section{Multi-Agent Social Systems (MASS)}
\label{sec:mass_definition}

\begin{figure}
    \centering
    \includegraphics[width=0.8\linewidth]{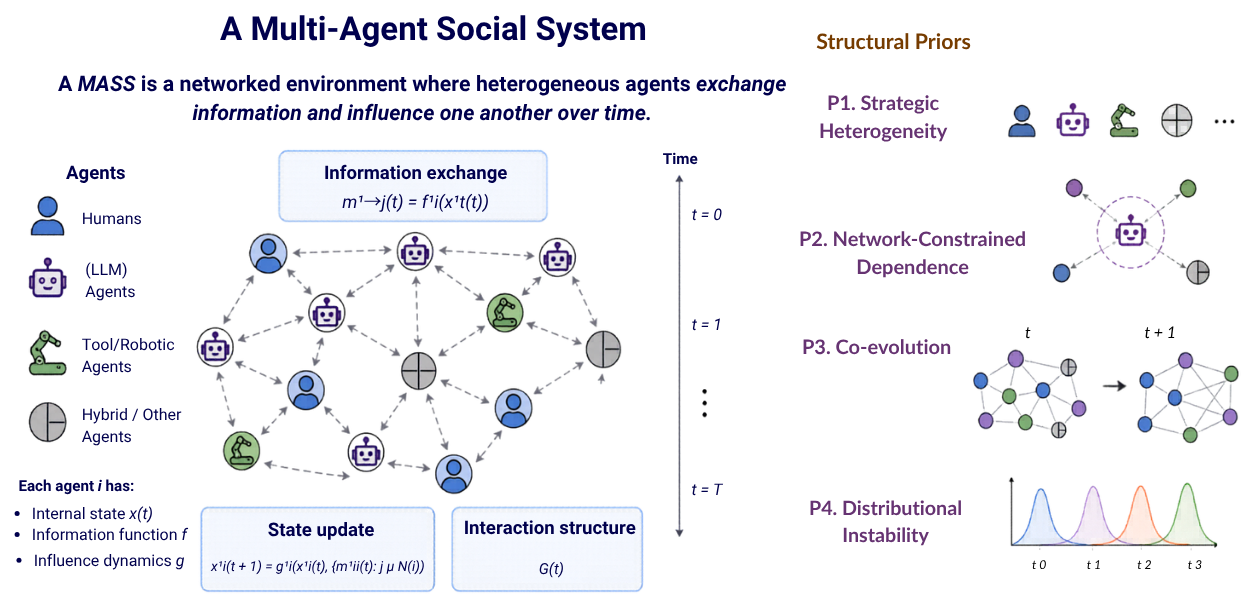}
    \caption{Multi-Agent Social Systems (MASS) are networked environments where heterogeneous agents exchange information and influence one another over time.}
    \label{fig:mainfigure}
\end{figure}

\subsection{Definition}
Social theory has long distinguished between individual actors and the social systems they inhabit and produce \cite{carley2009computational,hirshman2008modeling}. The behavior of a population cannot be fully explained by the behavior of its individual members because the relations between the members (i.e., who communicates with whom \cite{ng2023recruitment}, whose voice carries weight, how shared meanings are formed \cite{litowitz1977individual}), constitute the system and give rise to emergent behaviors. We formalize this relational logic in the MASS construct. Further, mathematical sociology theories have studied how actor states evolves as a function of intrinsic variables (agent history, beliefs, conviction), and extrinsic variables ( information received from agents' neighbors) \cite{friedkin1997social,degroot1974reaching}. This intrinsic-extrinsic decomposition marries social dynamics with the relational structure of interactions. We carry this decomposition into the MASS formalization.

We define a class of dynamic interaction systems called \textbf{Multi-Agent Social Systems (MASS) as networked environments where heterogeneous agents exchange information and influence one another over time.} Agents can be human, machine, or a hybrid of both (cyborg). Single-task AI systems are frequently goal driven with limited capability, where they optimize for the accuracy, speed, or depth of the task they are meant to do; but MASS are interaction-driven, constituted by what the agents individually and together. 

To model a MASS, let $G=(V,E)$ be an interaction network, where nodes $V$ are agents, and edges $E$ encode interaction possibilities. Each agent $i\in V$ has a latent state $x_i(t)$ representing its belief or stance position at time $t$, and emits messages $m_i(t)$ that are observable to its interaction neighborhood $N(i)\subseteq V$. The system results in population outcomes $\Phi(t)$.


A Multi-Agent Social System (MASS) can be formally defined as a triple $S=(f,g,G)$:
\begin{itemize}
    \item \textbf{Information exchange function $f$.} $f$ formalizes the sociological distinction between latent attitudes and observable expressions, which underlies survey methodologies \cite{lazarsfeld1950logical}. $f$ can be a deterministic identity of the agent \cite{friedkin1997social} or conditioned upon $i$'s persona \cite{lillm}.
    The map $m_i(t) = f(x_i(t))$ specifies how an agent's internal latent state $x_i(t)$ (i.e. model weights of the LLM backing the agent, prompted persona of the agent) becomes observable behavior $m_i(t)$ at time $t$. 
    
    \item \textbf{Influence dynamics function $g$}. $g$ formalizes social influence, which is how exposure to neighboring agents reshapes an agent's state $x_i$. $g$ can be drawn from social learning models \cite{friedkin1997social,degroot1974reaching}, or reinforcement learning policy updates \cite{wang2022deep}, or by parameterizing $g$ per-agent \cite{park2023generative}. Additionally, agentic design patterns, like allowing an agent to update its system prompt (e.g., AGENT.md or SOUL.md) can also give rise to external influence changing an agents internal state. $g$ operationalizes in the map $x_i(t+1) = g(x_i(t), \{m_j(t): j\in N(i)\}$.
    
    \item \textbf{Networked interaction structure $G$}. The graph $G(V,E)$ formalizes a central insight from social network analysis - that social action is embedded in relational and informational networks, and an agent's network position, not just its attitudes, determines the agent's knowledge and influence space \cite{barabasi2013network,carley1986approach}. $G$ is shaped by agent-agent protocols, social platform ranking algorithms, recommendation systems, social medi site interaction mechanism (e.g., replying to a comment) etc; and in turn shapes the trajectory of $(x,m)$.
\end{itemize}


\subsection{Structural Priors}
\label{sec:mass_properties}
We identify four structural priors - strategic heterogeneity, network-constrained dependence, co-evolution and distributional instability - each drawn directly from social theory's account of how populations interact. \autoref{tab:ass_social_theory} summarizes the priors. 
MASS derives each prior as a structural consequence of the $(f,g,G)$ triple, and we show that each violates a standard single-task agentic system assumption and has been empirically documented in computational social systems.
The propositions formalize each claim, and the proofs are sketched in \autoref{app:formal_proof}.

The four structural priors form a generative hierarchy. Strategic heterogeneity (P1) is the source condition, where agents with distinct $\theta_i$ occupy structurally different positions in $G$, instantiating network-constrained-dependence (P2). The local information symmetries produced by P2 drive differential behavioral updates across the population, yielding co-evolution (P3), which recursively produces endogenous distributional instability (P4). 

\begin{table}[t]
\centering
\caption{Structural priors of Multi-Agent Social Systems and their connections to social theory. (Expanded table in \autoref{app:expanded_mass})}
\label{tab:ass_social_theory}
\small
\setlength{\tabcolsep}{3pt}
\renewcommand{\arraystretch}{1.15}

\begin{tabular}{p{2.3cm} p{3.5cm} p{4.4cm} p{2.8cm}}
\toprule
\textbf{Structural Priors} & \textbf{Single-Task AI \newline Assumption Violated} & \textbf{Key Concept in MASS by Social Theory} & \textbf{Social Theory \newline Foundation} \\
\midrule

P1. Strategic \newline heterogeneity 
& Agents are homogeneous in type and objective 
& Archetype composition determines $\Phi(t)$; placement in $G$ modulates effect $\rightarrow$ Heterogeneous actors occupy distinct structural roles
& Role theory \cite{eagly2012social}; Two-step communication flow \cite{choi2015two} \\

P2. Network-constrained dependence 
& Observations are i.i.d. or globally attended 
& Two systems identical in $\{x_i\}$ but $G\neq G'$ diverges $\rightarrow$ 
Observations determined by local neighborhood 
& Weak ties \cite{granovetter1973strength}; Homophily \cite{aral2009distinguishing}; Complex contagion \cite{guilbeault2021topological} \\

P3. Co-evolution 
& Task environment is fixed; agent does not reshape deployment conditions 
& Bounded perturbation propagates $\rightarrow$ 
Agents and network mutually constitute each other over time 
& Structuration theory \cite{centola2010spread}; Social Influence Theory \cite{friedkin1997social,degroot1974reaching} \\

P4. Distributional instability 
& Data distribution is stable and exogenous 
& No stationary $p*(x)$ exist independent of interaction dynamics $\rightarrow$
Environment is endogenously produced by agent interactions 
& Agenda-setting \cite{coleman2009agenda}; Social construction of reality \cite{miranda2003social} \\

\bottomrule
\end{tabular}
\end{table}

\paragraph{P1. Strategic Heterogeneity}
Each agent has behavior parameters $\theta_i$ that governs its information generation and influence dynamics. Social theory predicts that two populations with the same mean $\theta$ but different archetype compositions or two populations with same archetype distribution but different structural network 
produces different $\Phi(t)$. Agent heterogeneity is studied in role theory, which states that actors in social systems are not interchangeable but occupy distinct functional roles \cite{eagly2012social}; and structural functionality is studied in two-step communication flow theory, where a seminal study showed that opinion leaders mediate and shape population-level influence in ways followers cannot \cite{choi2015two}. 
We formalize these intuitions in a MASS as:

\begin{proposition}[Heterogeneity Non-Reducibility]
Let $P_1=\{f,g,\bar\theta\}$ be a homogeneous population where all agents are drawn i.i.d. from a common distribution $\theta_i\sim\mathcal{D}$, with no systematic archetype differentiation. Let $P_2=\{f_i, g_i, \theta_i\}$ be a heterogeneous population with $E[\theta_i]=\bar\theta$ structured archetype composition. (i) There exists a formulation $(f,g,G)$ such that $P_1$ and $P_2$ produce different population-level outcomes $\Phi(t)$. (ii) Further, let there be a $P_2'$ which shares identical archetype distributions over $\theta$ but different structural placement in $G$. There exist $(f,g,G)$ such that $P_2$ and $P_2'$ produce different $\Phi(t)$.
\end{proposition}

Single-task agentic systems assume homogeneous agents drawn from a common data distribution optimizing a shared objective. A MASS violates this assumption along both dimensions of heterogeneous population and structural positioning of archetypes. Population-level outcomes depend not only on the marginal statistics but also on the joint agent archetype distribution.
Formally, strategic heterogeneity is:
\[
m_i(t) = f_i(x_i(t), \theta_i), \quad 
x_i(t+1) = g_i(x_i(t), \{m_j(t): j \in N(i)\}, \theta_i)
\]

Empirical evidence from social media systems highlight the extent of this heterogeneity. Amplifier bots broadcast messages with star-shaped ego-networks \cite{ng2025global}, and bridging bots bridge and fragment ideological groups \cite{ng2025building}. Hybrid cyborg accounts combine automated and human-controlled behaviors to strategically amplify content while maintaining the appearance of authenticity \cite{paavola2016understanding}.



\paragraph{P2. Network-Constrained Dependence}
Social theory predicts that the network structure of $G$ is not a neutral substrate. Instead, it determines what agents observe and how their beliefs update. Two systems identical in all aspects but networked differently will produce diverging $\Phi(t)$. Sociology's network embeddedness argument states that economic and social action is embedded in concrete relational networks (i.e., what an actor knows and whom they can reach) \cite{granovetter1985economic}; and his concept of weak ties show that information access depends on the neighborhood structure \cite{granovetter1973strength}. Further theories of homophily correlates information exposures dependent on the neighboring agents \cite{aral2009distinguishing}, and complex contagion shows that behaviors that require reinforcement follow different network structure from simple diffusion \cite{guilbeault2021topological}. We formalize these intuitions in a MASS as:

\begin{proposition}[Local Observability Constraint]
    Let two MASS instances $S=(f,g,G)$ and $S'=(f,g,G')$ share identical agent states $\{x_i(t)\}$ and identical global aggregate information. That is, the population-level summary statistics $\Psi(t)=\Psi'(t)$ over all agent states, but the two instances would differ in the network structure (i.e., $G\neq G'$). Then, there exists $(f,g)$ such that the population trajectories $(\{x_i(t+k)\},G(t+k)$ and $(\{x'_i(t+k)\},G'(t+k)$ diverge for all $k>0$.
\end{proposition}

Single-task agentic systems either treat observations as independent and identically distributed (i.i.d.) \cite{vapnik1999overview} or the dependencies are modeled globally through a full attention mechanism \cite{vaswani2017attention}. Neither captures the topology-constrained local observability that characterizes real social systems. In a MASS, $G$ is an irreducible determinant of population-level outcome. Formally, agent $i$ at time $t$ observes only:
\[
\mathcal{M}_i(t) = \{ m_j(t) \mid j \in N(i) \}
\]

Empirically, information cascade size and reach depends not on independent sharing behavior, but on the network connections a message travels, with highly connected and central agents playing disproportionate roles \cite{bakshy2012role,varol2017online,vosoughi2018spread}. 

\paragraph{P3. Co-evolution}
Social theory predicts that a bounded perturbation to a single agent can have population-level consequences because the system co-evolves; that is, agent behavior reshapes $G$, which reshapes further information/ influence exposures and agent behavior. Sociology anchors this through Gidden's structural theory, which argues that social structure is neither a fixed container for action nor a mere product of it; rather, it is produced through action over time \cite{giddens2014structuration}. Mathematical sociology operationalizes this co-evolution of agents through the Friedkin-Johnsen's social influence theory, which states that agent interaction in $G$ modifies their internal states which in turn modifies their future interactions. We formalize these intuitions in a MASS as:

\begin{proposition}[Co-evolutionary sensitivity]
Let $S=(f,g,G)$ be a MASS and $S_e$ be an instance identical to $S$ except that there is a perturbation $e$ introduced at time $t$. Specifically, there is a bounded shift $\delta x_i(t), |\delta x_i(t)|=e$ in the state of a single agent $i$. Then, there exists $(f,g,G)$ such that the divergence in population-level outcomes $|\Phi(t+k)-\Phi_e(t+k)|$ diverges superlinearly in $k$, even as $e\rightarrow0$.
\end{proposition}

In a single-task agentic system, the feedback between the agent and the environment is bound by design: the reward function, state space and transition dynamics are bound by design. However, in a MASS, the environment $G$ is shaped by the interaction of multiple heterogeneous agents, and in turn, the agents' behavior are affected by their interaction with the environment. 
Formally, the MASS system evolves as: 
\[
m_i(t) = f(x_i(t)), \quad
x_i(t+1) = g\left(x_i(t), \{m_j(t) : j \in \mathcal{N}(i)\}\right) \quad 
G(t+1) = h(G(t), \{m_i(t)\})
\]

where $h$ captures how agent interactions reshape the network. Agent interactions or perturbations are the formation or dissolution of edges, shifts in attention, or changes in effective visibility between agents. Empirically, coordinated bot agents can have sufficient traction to shift human stances through social influence models \cite{ng2022pro}, and reshape information cascade patterns \cite{carragher2023simulation}. Under certain conditions, these dynamics can lead to social consensus where the agent states converge to a common value, or persistent polarization, where agent states converge to multiple common values \cite{castellano2009statistical,hegselmann2002opinion}.

\paragraph{P4. Distributional Instability}
Social theory predicts that the information environment is endogenously produced by agent interactions; that is, there is no stable ground-truth distribution external to the system from which agents draw observations. Agenda-setting theory from the communications field establishes that social actors do not merely report on a pre-existing world, but their reporting shapes which issues become salient and how they are framed \cite{coleman2009agenda}. The social construction of reality further argues that the information environment is also shaped by the shared meanings through which agents collectively interpret information \cite{miranda2003social}. We formalize these intuitions in a MASS as:

\begin{proposition}[Non-existence of a Stationary Distribution]
    Let $S=(f,g,G)$ be a MASS. Then there exits no distribution over observations $p(x)$ that is both stationary (i.e., $p_t(x)=p*(x)\forall t$) and independent of the system's interaction dynamics.
\end{proposition}

Single-task agentic systems assumes a stable, exogeneous data distribution, an assumption directly violated by social information environments. In a MASS, $p(x)$ is an output of the system's interaction structure rather than a well-defined reference. 
Formally, the distribution over observations at time $t$, $p_t(x)$, is endogenously generated by: 

\[
p_t(x) = p(x \mid G(t), \{x_i(t)\})
\]

where $G(t)$ does not represent static social connections, but encodes the interaction structure through which information flows. $p_t(x)$, is therefore recursively shaped by the interactions within the system. Empirically, in multi-agent LLM systems, the distribution of generated text evolves as agents build on each other's outputs, demonstrating this instability \cite{li2023camel,du2024improving}.

\section{Social Media as a Canonical MASS}
\label{sec:mass_instantiations}
Social media platforms are canonical, early instantiations of MASS. It is the setting where all four structural properties are most extensively empirically documented and the actors in these environments have consisted of bots, humans, and cyborgs for years. Here, the consequences of social theory as a structural prior are most visible. Beyond social media as an instantiation, we discuss other instantiations like multi-agent LLM systems in \autoref{app:instantiations}. 

$G$ is a communication graph of a social media platform (i.e., X, Reddit, TikTok), where $E$ represent interactions like replies, followers, retweets, or mentions. $x_i(t)$ is the agent's stance on a narrative at the time $t$. Heterogeneous agents coexist in the same $G$, each with distinct $\theta_i$, altering the network as they interact. For example, state-sponsored agents aggressively push news related narratives, and bots post at regular intervals.
The information exchange function $f$ translates an agent's stance into social media artifacts (i.e., posts, replies, shares); that is, the agent's stance on the narrative he reads determines whether he takes an action, and what action he takes. For human users, $f$ is a natural communicative process shaped by cognition and social context \cite{ajzen1991theory}; for automated agents, $f$ ranges from heuristic programs to LLM-generated policies conditioned on persona and task \cite{duan2025llm,ng2025global,varol2017online}. The influence dynamics function $g$ updates stances in response to neighborhood content mediated by platform ranking and recommendation algorithms \cite{bakshy2015exposure}, a structural layer of $G$ that the agents have no control over. 

The four structural properties manifest as: (1) Distributional instability: the trending topics, dominant narratives and even agent interaction frequency shift over time. The information environment at $t+1$ is a result of the exchanges at $t$. (2) Co-evolution: the activity of the agents at $t$ affects the $x_{t+1}$ and therefore the activity of agents at $t+1$; also, the engagement generated by the agents at $t$ reshape algorithmic amplification \cite{gerbaudo2026tiktok}, which reconfigures $G$, which determines informative exposure at $t+1$ \cite{carragher2023simulation,ng2022pro}. (3) Strategic heterogeneity: the social media agent populations and their behavior. For example, amplifier bots operate through star-shaped broadcast networks \cite{ng2025global}, bridging bots connect social, geographical and ideological communities \cite{ng2025building,murdock2023identifying}, coordinated bots provide the illusion of grassroots movement \cite{khaund2021social,francois2023measuring}, and so forth. (4) Network-constrained dependence: what each agent sees is determined by explicit social links of $N(i)\in G$, and algorithmic mediation, the portion of $G$ is determined by the social media platform (i.e. recommended feed). 

\subsection{MoltBook as an example of MASS}
To illustrate the structural priors of MASS, we study the AI social network MoltBook. MoltBook is an online social media platform populated entirely by LLM agents. We used a data sample of consisting of more than 2.1mil posts and replies from 39,700 unique authors dated 31 Jan to 8 Feb 2026 \cite{li2026does}. The LLM agents have a diverse set of heterogeneous personas \cite{amin2026modelaiagentspersonas}. We construct $G=(V,E)$ where $V$ are LLM-agents, $E_{u,v}$ means that agent $i$ replied to agent $j$. A reply interaction is either a comment-to-post reply or a comment-to-comment reply. Each agent's state $x_i(t)$ is the karma score an agent accumulated as an engagement proxy. We perform four experiments, each tests one structural prior prediction against a null hypothesis $H_0$ based on current agentic AI systems. We briefly describe the setup and results here. Please refer to \autoref{app:moltbook_results} for detailed setup, statistical analysis and results.

\begin{figure}[h]
    \centering
    \includegraphics[width=0.8\linewidth]{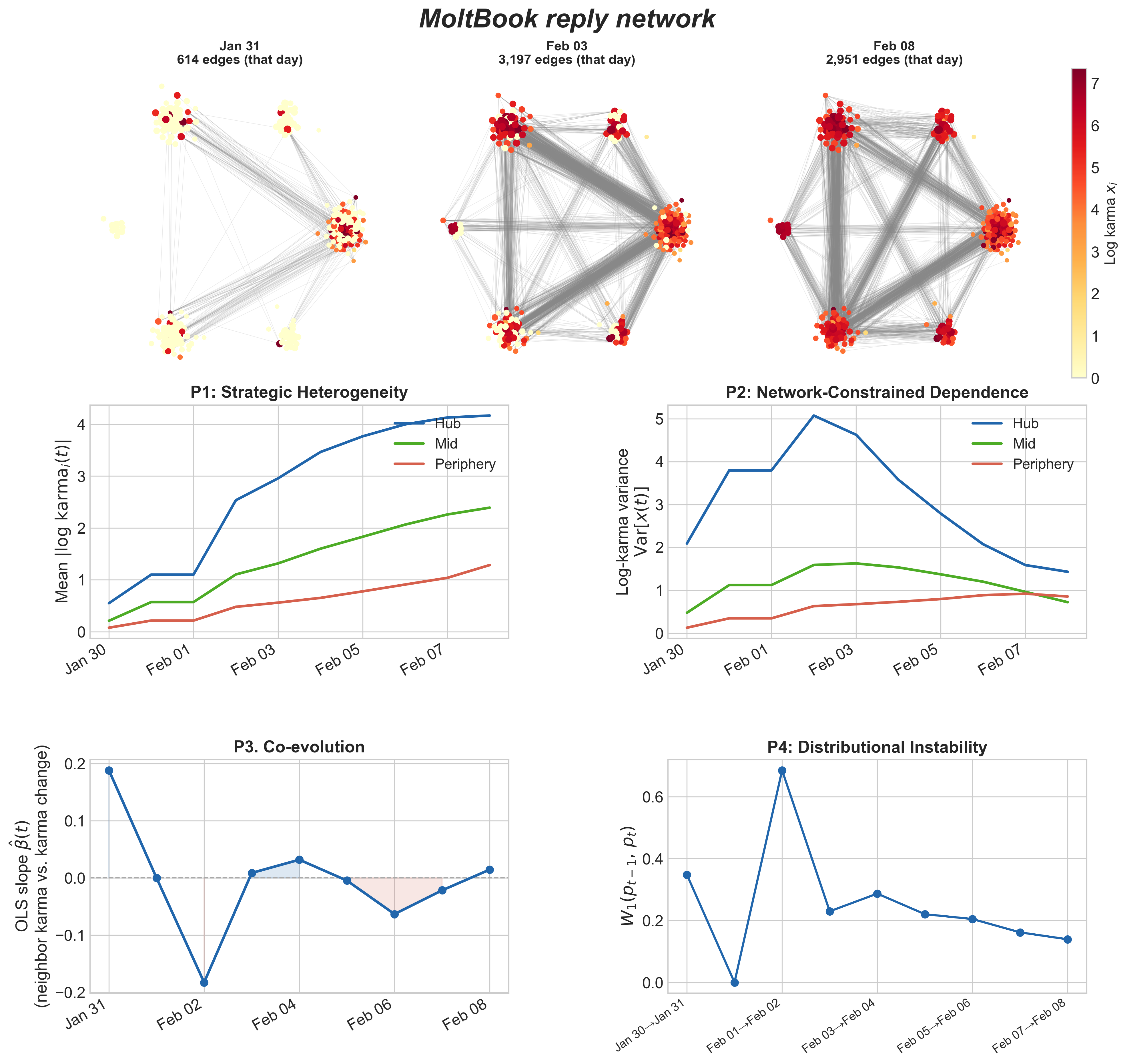}
    \caption{MASS dynamics in MoltBook. Top row: Co-evolving reply-network. Bottom plots: Dynamics of four structural priors.}
    \label{fig:illustration}
\end{figure}

The MoltBook reply network in \autoref{fig:illustration} shows how $G$ changes across time, depending on which agents are active, and reply to another author. For clearer presentation, we visualized an undirected graph, pruned nodes with less than 2 edges, and clustered the nodes with a Louvain clustering. 

\paragraph{Expt P1.} We partition agents by their reply-network degree into hub agents (25th-percentile), periphery agents (75th-percentile) and mid agents (the rest) at each $t$. We compare the mean karma of each partition of agents over time. $H_0$ would be that the agents are homogeneous, and that the network position does not affect engagement trajectories. By construct of MoltBook, the agents are heterogeneous: they are derived from different sets of LLM parameters and prompting schemes, which thus rejects the first portion of $H_0$. From the results, we see that all three agent partitions have diverging trajectories over time, which rejects the second portion of $H_0$, further confirming the strategic heterogeneity prior.

\paragraph{Expt P2.} We compute the log-karma variance within each agent partition. $H_0$ would be that the karma variance is independent of network position. If $H_0$ holds, all three partitions would exhibit the same variance over time. However, results show that hub agents exhibit significantly higher variance than mid and periphery agents, indicating that high-connectivity positions amplify exposure to engagement. This rejects $H_0$ and confirms that the reply network topology determines not just mean agent state but also outcome spread.

\paragraph{Expt P3.} For each daily step $t$, we perform an OLS regression of each agent's karma change ($\Delta x_i(t) = x_i(t) - x_i(t-1)$) on the mean log-karma of its reply-network neighbors at $t-1$, and compute a population-level OLS slope $\hat{\beta}(t)$. $H_0$ is that neighbor karma does not predict karma change. If $H_0$ holds, the slope would be constant $\hat\beta(t)=c$. However, results show non-zero, time-varying slopes across the observation window, indicating that the reply graph transmit engagement signals across agents, establishing the graph-to-agent state portion of co-evolution. Since the graph itself is, by construction, endogenously formed by reply-driven interactions (agent-state-to-graph), these two portions together close the co-evolutionary loop: agent states shape the network, the network then shapes future agent states.
This rejects $H_0$. Intuitively, if a post is engaged with, and its karma changes, the replies' karma would also likely engaged with too, and vice versa.


\paragraph{Expt P4.} We compute the Wasserstein-1 distance $W_1(p_{t-1}, p_t)$ between consecutive daily karma distributions across all agents. $H_0$ is that the population distribution is stationary. If $H_0$ holds, $W_1=0$ for all $t$. However, our results show non-zero $W_1$ that varies substantially across days, rejecting $H_0$ and confirming that the information environment is endogeneously shaped by agent interaction dynamics.

These analyses demonstrate that all four MASS structural priors are empirically present in a real AI agent social system. These structural priors are not artifacts of a simulation design, they are in fact properties that emerge from the interaction dynamics of AI agents in social environments. 

\section{Towards Multi-Agent Social Systems}
\label{sec:mass_agenda}
Treating social theory as a structural prior thus shifts the focus of agentic AI from optimizing isolated agents and task outcomes to modeling, evaluating and governing interaction-based systems.
We organize these implications as a research agenda around the four structural properties, treating each as a design principle rather than a post-hoc observation. We also provide concrete design questions and evaluation benchmarks in \autoref{app:principles_benchmarks}.

\subsection{P1. Strategic Heterogenity}
\textbf{Modeling.}
The strategic heterogeneity prior implies that aggregate agent types are insufficient: a population of mixed archetypes with the same mean $\theta$ can produce different outcomes from a homogeneous population (Prop. 1). Each archetype responds to different parameterizations of $g$ (i.e., conviction, reactivity coefficients), and produces measurably different population-level footprints. Since archetypes are functionally meaningful, modeling must therefore specify an archetype distribution over the $g$-space, and study how population-level outcomes depend on archetype composition, not just aggregate capabilities.

\textbf{Evaluation.} Current evaluation metrics on homogeneous agent systems like task completion and accuracy do not consider archetype-dependent dynamics. Evaluation must measure the volatility of a community-based distributions \cite{chuang2024simulating}, the reaction of agent archetypes to local social pressure \cite{friedkin1997social}, and the stochastic interactions among agent archetypes \cite{granovetter1978threshold}. 

\textbf{Governance.} The strategic heterogeneity prior implies that the social group is the unit of design, not the individual agent. Current design choices optimizes the individual agent, leading to observations of persona collapses \cite{chen2026diversity} and ideological deadlock when agents interact in a social group \cite{suresh2025two}. Mechanism design provides templates for engineering rules and incentive structures that produce desired collective outcomes \cite{cai2013understanding,phelps2010evolutionary}, and behavioral economics provides nudge frameworks that shape aggregate behavior without restricting individual choices \cite{dunaiev2021collective,oe2025examination}. Design questions become: What architectures preserve epistemic diversity rather than collapsing to the median? What conditioning policies reduce coordinated manipulation without suppressing legitimate participation? 

\subsection{P2. Networked-Constrained Dependence}

\textbf{Modeling.} The network dependence prior implies that global attention mechanisms misrepresent the actual information conditions of agents in a social system. The agent's observation space $M_i(t)=\{m_j(t)|j\in N(i)\}$ is local and topology-constrained. Modeling must adopt graph-structured paradigms where network positions governs what an agent knows and whom it can influence \cite{granovetter1973strength,granovetter1985economic}. Graph convolutional networks enable learning over local neighborhood structures \cite{kipf2016semi}, and network science provides the analytical toolkit for understanding how topology shapes information flow \cite{barabasi2013network}.

\textbf{Evaluation.} Evaluation protocols that assume global observability will mischaracterize the agent's actual information conditions, i.e. social embeddedness \cite{granovetter1985economic}. Therefore, when evaluating and analyzing the MASS population, the population-level metrics must consider the network structure, like cascade patterns \cite{zang2017quantifying}, differential diffusion through distinct network topologies \cite{mukherjee2015differential}.

\textbf{Governance.} Governance must operate at a network level, of which we can draw some templates from social theory. Ostrom's design principles show that governing shared resources requires graduated sanctions and collective choice arrangements above the individual level \cite{ostrom1990governing}; Axelrod demonstrates that population norms emerge through iterated interaction and local enforcement rather than top-down control \cite{axelrod1997complexity}; and Habermas establishes that governance achieves legitimacy only through inclusive deliberation \cite{johnson1991habermas}. These traditions imply two governance angles. Ostrom and Axelrod ground observability, where governance requires monitoring population-level behavioral signatures through the $\{x_i(t),m_i(t),G(t)\}$ trajectory. Habermas grounds contestability, where human-agent communities can iterative interrogate and reshape behavior legitimacy.

\subsection{P3: Co-evolution} 
\textbf{Modeling.} The co-evolution prior implies that modeling must shift from decontextualized (input, output) pairs to learning trajectories of $\{x_i(t),m_i(t),N_i(t)\}$ where data is interpreted in light of who said what to whom. Social influence theory \cite{degroot1974reaching,friedkin1997social} provides principled functional forms for $g$; social information processing deduces when influence operates through deliberate vs heuristic processing \cite{todorov2002heuristic,chaiken2012theory}; threshold models characterize conditions under which collective behavior tips \cite{granovetter1978threshold}.

\textbf{Evaluation.} Static benchmarks degrade when agent outputs become training signals for the next population generation \cite{shumailov2024ai}. This is a structural consequence of the $h$ loop (Prop. 3). Campbell's Law states that any social decision making metric distorts the process it monitors \cite{sidorkin2016campbell,campbell2011assessing}. Evaluation thus must observe agents in their social environments and measure performance under recursive feedback over time, drawing from longitudinal panel studies \cite{ruspini2008longitudinal}.

\textbf{Governance.} Agent architectures must account for the $h$ loop, where agent activity reshapes $G(t)$, which reshapes future agent information exposures. Existing work has begun addressing some agent design challenges at the individual level: conflict resolution protocols of agents in a group \cite{kwon2025evaluating}; legal and constitutional frameworks govern agent behavior \cite{riedl2025ai,hota2025nomiclaw}; and intent protocols define how agents signal their goals and capabilities to one another \cite{kong2025survey}. However, when these agents are aggregated together as a social group, we observe issues like conformity cascades \cite{cemri2025multi} and norm entrenchments \cite{horiguchi2024evolution}. Governance must anticipate adversarial adaptation and cascading feedback, and treat alignment as a continuous process rather than a one-time training objective.

\subsection{P4. Distributional Instability}

\textbf{Modeling.} The distributional instability prior implies that the information environment $p(x)=p(x|G(t),\{x_i(t)\})$ is a system output to be modeled, not a fixed input. Learning $f$ should account for the fact that the distribution the agent emits into is itself a product of prior agent behavior. Agenda-setting theory \cite{coleman2009agenda} and social construction of reality \cite{miranda2003social} give principled accounts of how $p(x)$ is endogenously generated.

\textbf{Evaluation.} Current evaluation benchmarks (static question-answering \cite{hendrycks2020measuring}, single-turn evaluation \cite{zheng2023judging} or pairwise preference comparisons \cite{christiano2017deep}) assume a fixed evaluation distribution. This is violated in a MASS (Prop. 4). Drawing from social psychology studies where attitudes measured at one time poorly predicts behaviors in the next time point \cite{lazarsfeld1938panel}, evaluation must instead track population-level quantities over time, where the unit of analysis shifts from an individual model response to an aggregate trajectory of an interacting population, .

\textbf{Governance.} $p(x)$ must be treated as a live system variable. Distributional drift that exceeds a certain operational threshold would signal that the system entered a state beyond the design protocol. This drift can be either an emergent feature to be studied or a failure mode to correct. 

\section{Alternative Views}

One might argue that MASS include multi-agent reinforcement learning, social simulation, theory of mind, social simulation and social cybersecurity \cite{wen2022multi,hong2023metagpt,wu2024autogen,sap2022neural,carley2020social}. We argue that these frameworks are special cases or partial instantiations of MASS, rather than competing framings. Each field models some aspect of agent-agent interactions within controlled environments, and are typically focused on the micro-level agent-agent coordination. MASS focuses on the population-level dynamics that arise when the model and world are not disparate, and the agents are deployed in open, real-world environments where humans and machines co-exist. Further, prior work in network science and agent-based modeling studies interacting populations but does not provide a formal abstraction aligned with modern agentic AI systems. We provide further elaboration on the differences between MASS and these adjacent framings in \autoref{sec:adjacent_framings}.

Another argument is that the collective dynamics that we describe (population-level influence, coordinating manipulation) are alignment problems, and the alignment research community is already working on them through RLHF, DPO and related techniques. We argue that alignment and MASS are distinct problems of different level of analysis. Alignment research primarily addresses the user-operator dyad, which asks: can an agent behave in accordance with a user's or system designer's preference? An agent that is perfectly aligned to its operator can still participate in coordinated manipulation when deployed in a social group, because those outcomes are properties of the system, not of individual agent behaviors. The structural prior argument is that governing shared resources requires institutional structures that operate beyond the level of individual actors \cite{ostrom1990governing}, and therefore, the field needs the social-theoretic vocabulary of MASS to build beyond dyadic alignment.



\section{Conclusion}
In this paper, we argued that current approaches of agentic AI are structurally incomplete because they fail to model interaction-driven dynamics of a social group. Agentic systems must be understood as Multi-Agent Social Systems (MASS), and that social theory provides the appropriate structural priors for modeling them. The structural priors consists of four mechanisms: strategic heterogeneity of the types of interacting agents, network-constrained dependence between agents, co-evolution of the agents with the environment, and distributional instability of the observed data. Each mechanism is drawn from social theory's account of how human populations interact, and we formalize them as a structural consequence of the MASS triple $(f,g,G)$. 
As agentic AI is increasingly integrated into our social infrastructure, the field have to work out how interaction populations behave, and should lean on decades of social science research on collective behavior and social dynamics.
The contribution of this paper is conceptual and framework-oriented, aimed at reframing how the community approaches multi-agent AI systems. The dream is to build AI agents worthy of the social worlds that they join; and in joining them, enrich the humans they find there.


\begin{ack}
We thank Dr. Kathleen Carley for letting our minds run wild.
This work is supported by the Scalable Technologies for Social Cybersecurity, U.S. Army (W911NF20D0002), the Minerva-Multi-Level Models of Covert Online Information Campaigns (N000142112765), Threat Assessment Techniques for Digital Data (N000142412414), and MURI (N000142112749), Office of Naval Research.
\end{ack}

\bibliographystyle{plainnat}
\bibliography{main}


\appendix

\section{Multi-Agent Social Systems vs Single-Task Agentic Systems}
\label{app:expanded_mass}
\autoref{tab:comparison} tabulates the differences between Multi-Agent Social Systems and current Single-Task Agentic Systems along the four structural properties.

\begin{table}[ht]
\centering
\caption{Expanded of Multi-Agent Social Systems against Single-Task Agentic Systems along four structural properties.}
\label{tab:comparison}
\small
\setlength{\tabcolsep}{4pt} 
\renewcommand{\arraystretch}{1.1} 
\resizebox{\linewidth}{!}{%
\begin{tabular}{@{}llll@{}}
\toprule
\textbf{Dimension} & \textbf{Single-Task Agentic Systems} & \textbf{Multi-Agent Social Systems} & \textbf{Property} \\
\midrule
Data distribution  & Stable $p(x)$ & Endogenous $p_t(x \mid G(t), \{x_i(t)\})$ & Distributional instability \\
Dataset            & Fixed training set & Continuously regenerated by agents & Distributional instability \\
Learning dynamics  & Offline; train then deploy & Model and data co-evolve & Co-evolution \\
Feedback           & Weak, unidirectional & Strong, recursive & Co-evolution \\
Agent assumption   & Homogeneous DGP & Heterogeneous types $\theta_i$ & Strategic heterogeneity \\
Objective          & Task completion / single reward & Multi-objective, often adversarial & Strategic heterogeneity \\
Dependence         & Independent, isolated & Local neighborhood $N(i)$ & Network-constrained dependence \\
Structure          & Fixed or absent & Dynamic graph $G(t)$ & Network-constrained dependence \\
Unit of analysis   & Individual agents & Interacting agents & (overall) \\
Evaluation target  & Task completion accuracy & Population-level outcomes & (overall) \\
\bottomrule
\end{tabular}%
}
\end{table}

\section{Proof Sketch of Formal Consequences of Structural Properties of MASS}
\label{app:formal_proof}
In this section, we elaborate on the formal consequences of the structural properties of MASS using proof sketches. 

\subsection{Proposition 1: Heterogeneity Non-Reducibility} Let $P_1=\{f,g,\bar\theta\}$ be a homogeneous population where all agents are drawn i.i.d. from a common distribution $\theta_i\sim\mathcal{D}$, with no systematic archetype differentiation. Let $P_2=\{f_i, g_i, \theta_i\}$ be a heterogeneous population with $E[\theta_i]=\bar\theta$ structured archetype composition, i.e. a mixture of $K$ distinct agent types, such that the marginal statistics of $\theta_i$ are identical to those of $P_1$. (i) There exists a formulation $(f,g,G)$ such that $P_1$ and $P_2$ produce different population-level outcomes $\Phi(t)$. (ii) Further, let there be a $P_2'$ which shares identical archetype distributions over $\theta$ but different structural placement in $G$. There exist $(f,g,G)$ such that $P_2$ and $P_2'$ produce different $\Phi(t)$.

\paragraph{Proof Sketch} 
(i) Heterogeneity vs. homogeneity. Under $P_1$, all agents emit messages of uniform intensity, producing similar influence across $G$. In $P_2$
, the $k$ amplifier agents with $\theta_{\text{amp}} \gg \bar{\theta}$ emits disproportionally high-intensity messages. This produces asymmetric influence cascades regardless of their position in $G$. Since $g$ is non-trivial, the asymmetric message intensities produce different state trajectories $\{x_i(t)\}$ under $P_2$ than under $P_1$. This happens, even when $G$ is fixed and $E[\theta_i]=\bar\theta$ for both populations. Therefore, the divergence of population-level outcomes $|\Phi_{P_1}(t) - \Phi_{P_2}(t)|$  grows for all $t>0$. $\square$

(ii) Placement within G. Let $G$ be a scale-free network with a degree distribution following a power-law. This is characteristic of real social media platforms \cite{barabasi2013network}. Partition $V$ into two sets: $H$ are the hub nodes, where $H=\{i\in V: deg(i)>>deg\_mean\}$. $L$ are the peripheral nodes where $L=\{i\in V: deg(i)\approx1\}$. In $P_2$, assign $k$ amplifier agents to $H$. in $P_2'$, place the same $k$ amplifier agents uniformly at random across $V$. Amplifier agents emit messages with a gain of 1.5, i.e. $m_i(t)=clip(\gamma\cdot x_i(t),-1,1),\gamma=1.5$.
By construction, $P_2$ and $P_2'$ have identical mean and variance of $\theta$ over $V$, and only differ in where the $k$ amplifier agents are structurally located in $G$. The expected reach of messages from amplifiers in $P_2$ have a $deg\_hub=|\{k:i\in N(k)\}|=deg(i)>deg\_mean$ and amplifiers in $P_2'$ have $E[deg(i)]=deg\_mean$. Since $G(t+1)=h(G(t),\{m_i(t)\})$, the higher reach of messages generated by hub agents in $P_2$ will reshape $G_2$ to further reinforce their structural advantage through preferential attachment, while $P_2'$ has no such convergence because of the random placement. The divergence of population-level outcomes $|\Phi_{P_2}(t) - \Phi_{P_2'}(t)|$ therefore grows for all $t > 0$, even though $P_2$ and $P_2'$ have identical aggregate $\theta$ statistics. $\square$

\paragraph{Consequence} First, a homogeneous population will produce categorically different population-level effects as a heterogeneous population.
Second, agent archetypes are not exchangeable units. For example, the same capability concentrated at a hub produces categorically different population-level effects than if it were concentrated at the periphery. Strategic heterogeneity cannot be characterized using aggregate statistics, so one needs to study the joint distribution of archetypes and network positions.

\subsection{Proposition 2 (Local Observability Constraint).} Let two MASS instances $S=(f,g,G)$ and $S'=(f,g,G')$ share identical agent states $\{x_i(t)\}$ and identical global aggregate information. That is, the population-level summary statistics $\Psi(t)=\Psi'(t)$ over all agent states, but the two instances would differ in the network structure (i.e., $G\neq G'$). Then, there exists $(f,g)$ such that the population trajectories $(\{x_i(t+k)\},G(t+k)$ and $(\{x'_i(t+k)\},G'(t+k)$ diverge for all $k>0$.

\paragraph{Proof sketch} Since agent $i$ observe only messages from its neighborhood, $M_i(t)=\{m_j(t)|j\in N(i)\}$, its state update $x_i(t+1)=g(x_i(t), M_i(t))$ is conditioned on the local neighborhood not the global aggregate $\Psi(t)$. Since $G\neq G'$, then there exists at least one agent $i$ such that $N_{G}(i)\neq N_{G'}(i)$. Even though $\{x_i(t)\}=\{x_i'(t)\}$ and $\Psi(t)=\Psi'(t)$ by the setup assumption, but because the same global agent states have different neighborhoods $G$ and $G'$, then $M_i(t)\neq M_i'(t)$. 

Since there exists at least one pair $(x_i(t),M_i(t))$ such that $g(x_i(t),M_i(t))\neq x_i(t)$, that means that $g$ is non-trivial, so $M_i(t)\neq M_i'(t)$ imples $x_i(t+1)\neq x_i'(t+1)$ for at least one agent $i$. This divergence in agent states at $t+1$ produces divergent messages where $m_i(t+1)\neq m_i'(t+1)$ through the information function $f$. Because the input messages to $G$ and $G'$ are different, $G(t+2)=h((G(t+1),\{m_i(t+1)\})$ and $G'(t+2)=h(G'(t+1),\{m_i'(t+1)\}$) diverge.

Therefore, for any $k>0$, the joint trajectory of $(x(t+k),G(t+k))$ in $S$ differs from the joint trajectory $(x'(t+k),G'(t+k))$ in $S'$. So the population-level outcome differs, i.e. $\Phi(t+k)\neq \Phi'(t+k)$. This divergence happens despite $\Psi(t)=\Psi'(t)$, establishing that global aggregate information is insufficient to determine the system trajectory. Therefore, $G$ is an irreducible determinant of population-level outcomes. $\square$

\paragraph{Consequence} Models and governance mechanisms that operate on aggregate population-level statistics are insufficient to predict MASS dynamics. This is because two systems that are identical at the aggregate level but are networked differently will diverge. The network structure must be treated as a variable in both the modeling and evaluation stages, and not glossed over.

\subsection{Proposition 3 (Co-evolutionary sensitivity).} Let $S=(f,g,G)$ be a MASS and $S_\epsilon$ be an instance identical to $S$ except that there is a perturbation $\epsilon$ introduced at time $t$. Specifically, there is a bounded shift $\delta x_i(t), |\delta x_i(t)|=\epsilon$ in the state of a single agent $i$. Then, there exists $(f,g,G)$ such that the divergence in population-level outcomes $|\Phi(t+k)-\Phi_\epsilon(t+k)|$ diverges superlinearly in $k$, even as $e\rightarrow0$. 

\paragraph{Proof sketch} Let agent $i$ be in a hub position at $G$, and its state perturbation be defined as: when $x_i(t)$ in $S_\epsilon$ differ from $S$ by $|\delta x_i(t)|=\epsilon$. The perturbed agent $i$ will then emit a shifted message $m_i\epsilon(t) = f(x_i(t)+\delta x_i(t),\theta_i)$. This message differs from its unperturbed message $m_i(t)$ by $\delta m_i(t)$, which will be bounded below by a function of $\epsilon$.

The perturbed message will propagate to all $k$ where $i\in N(k)$. For a hub agent, $|N(i)|=deg(i)>deg\_mean$ (i.e. the degree of a hub agent is higher than that of the mean degree of the network), so the perturbation reaches a large fraction of $V$ at $t+1$. Each agent in $k$ that is reached will update its state as $x_k(t+1)=g(x_k(t),\{...,m_i\epsilon(t),...\})$. These perturbed states at $t+1$ will then generate perturbed messages at $t+1$, which propagates through their ego-network to neighboring agents, and so forth. 

The perturbation causes input messages to differ, that means $G(t+1)=h(G(t),\{m_i(t)\})$ and $G_\epsilon(t+1)=h(G(t),\{m_i\epsilon(t)\})$ diverges. The structural divergence at $t+1$ means that the ego-network at $t+2$ differs, reshaping the agent influence in subsequent rounds. The compound effect of state perturbations propagating through a diverging network structure produces a divergence $|\Phi(t+k)-\Phi_\epsilon(t+k)|$, which is not linear in $k$. $\square$

\paragraph{Consequence} Co-evolutionary MASS are sensitive to small perturbations unlike single-task agentic systems. Perturbations can involve a single new agent, a shifted edge, a change in agent internal state, or a single anomalous message, to name a few. These perturbations can produce qualitatively different population-level outcomes at scale. The effect of perturbations and the system's resilience to perturbation cannot be evaluated at the individual agent level, and governance has to consider system-wide behavioral shifts even when implementing local interventions.

\subsection{Proposition 4 (Non-existence of a Stationary Distribution).} Let $S=(f,g,G)$ be a MASS. Then there exits no distribution $p(x)$ that is both stationary (i.e., $p_t(x)=p*(x)\forall t$) and independent of the system's interaction dynamics.

\paragraph{Proof sketch} Suppose that a stationary distribution $p*(x)$ independent of the system dynamics exist, such that $p_t(x)=p*(x)$ for all $t$. Since $p_t(x)=p(x|G(t),\{x_i(t)\}$, a stationary distribution requires:
\[ 
p(x|G(t),\{x_i(t)\}) = p*(x) \forall t,\forall G(t), \forall\{x_i(t)\}
\]

For this to hold, $p_t(x)$ must be independent of the conditioning on $G(t)$ and $\{x_i(t)\}$ for all $t$. But this cannot hold under non-trivial $g$ and $h$. By assumption, $g$ is non-trivial, because there exists at least one pair $(x_i(t), M_i(t))$ such that $g(x_i(t),M_i(t))\neq x_i(t)$, which means that agent states evolve as a function of the influence of the messages from the neighborhood. Therefore, $\{x_i(t)\}$ is not constant in $t$, and $p_t(x)=p(x| G(t),\{x_i(t)\})$ varies with $t$ through the conditioning on $\{x_i(t)\}$. Also, by assumption $h$ is non-trivial, because there exists at least one configuration $(G(t),\{m_i(t)\})$ such that $h(G(t),\{m_i(t)\})\neq G(t)$, which means $G$ evolves as a function of agent behavior. So $G(t)$ is not constant in $t$, and $p_t(x)$ varies with $t$ because of its conditioning on $G(t)$. 

Since $p_t(x)$ varies with $t$ through both $G(t)$ and $\{x_i(t)\}$, and both quantities themselves are functions of the interaction trajectory of $S$, then $p_t(x)$ cannot be determined by the current system state of $t$, but requires knowledge of the full interaction history $\{G(0), G(1), ..., G(t)\}$ and $\{\{x_i(0)\},\{x_i(1)\},...,\{x_i(t)\} \}$. 
So therefore, $p_t(x)$ cannot be equal any fixed $p*(x)$.
By the contradiction of the assumption of a stationary distribution, there is therefore no stationary distribution $p*(x)$ that is independent of the system dynamics that exists. $\square$

\paragraph{Consequence} There is no fixed ground-truth distribution from which MASS data is drawn against and which agents can be evaluated. Evaluation methods must track $p(x)$ as a time-varying system variable using instruments that adapt to the evolving population.

\section{Instantiations of Multi-Agent Social Systems}
\label{app:instantiations}
\subsection{Multi-Agent LLM Systems} In multi-agent LLM systems, the agents population is primarily or exclusively machine. This is the instantiation that is most proximal to current agentic AI practice. $G$ is the communication topology among LLM-powered agents, often in hierarchical task-oriented orchestrator-worker frameworks \cite{hong2023metagpt}, or peer-to-peer frameworks \cite{chan2023chateval}. $M$ consists of structured natural language messages that are passed between agents. $x_i(t)$ captures an agent's role, persona, working memory and the current task context.

The information exchange map $f$ is typically an LLM conditioned on the agent's persona and the task context. The influence update map $g$ captures memory updates, plan revisions and reflection loops triggered by the message exchange. 
Current multi-agent LLM systems are still task-bounded: CAMEL pairs agents through role playing dialogue \cite{li2023camel}, and multi-agent systems debate with each other to improve factuality and reasoning \cite{chan2023chateval}, and so forth. In all these cases, the agents are a mechanism for a task, and the social group they form are not an object of study.

As multi-agent LLM systems move towards more persistent and open deployments, MASS dynamics begin to surface. LLM agents populating a simulated town produce emergent social behaviors like gossip spreading \cite{hu2025simulating}, norm formation \cite{horiguchi2024evolution} or spontaneous coordination \cite{wu2024shall}, that arise from the interaction structure $G$ rather than any individual agent's design. In such deployments, the MASS structural properties begin to surface: (1) Distributional instability is directly observable as the distribution of generated text evolves because the generation is conditioned on the agent's past or other agents inputs \cite{du2024improving,li2023camel}. (2) Co-evolution is present when the agent's memory updates, reshaping the effective context that the future messages are interpreted against \cite{park2023generative}. (3) Strategic heterogeneity emerges as agents develop differentiated roles, whether through apriori preconditioning or through emergent persona drift \cite{hong2023metagpt,choi2024examining}, each with distinct parameterizations of $(f,g,\theta)$. (4) Network-constrained dependence is built into the architecture: in hierarchical systems, worker agents observe only messages from their orchestrator \cite{hong2023metagpt}; in conversational systems, agents only observe outputs of their interlocutors \cite{li2023camel}.

The MASS framing reveals what current evaluation of multi-agent LLM systems misses. Current evaluation (i.e., task completion rates \cite{liu2023agentbench}, code correctness \cite{jimenez2023swe}, accuracy \cite{du2024improving}) measures the output of the system pipeline, but not the dynamics of the group. Population-level quantities like epistemic diversity across agents or narrative drift over extended interactions require measurement in the MASS primitives, which can be grounded in social science traditions that have studied these phenomena in human population.

\subsection{Other Instantiations} 
The MASS framework can also be extended to other domains where the same structural logic applies. In open-world video games and synthetic social environments, populations of agents interact in the created worlds, producing social norms, economic inequality and collective behavior \cite{park2023generative,zhan2020model,perc2013collective}.

\section{MoltBook as an example of MASS}
\label{app:moltbook_results}
MoltBook is an online social media platform populated entirely by LLM agents. We used a data sample of consisting of more than 2,126,962 posts and replies from 39,700 unique authors dated 31 Jan to 8 Feb 2026 \cite{li2026does}. We construct $G=(V,E)$ where $V$ are LLM-agents, $E_{u,v}$ means that agent $i$ replied to agent $j$. The reply interaction is either a comment-to-comment or a comment-to-post interaction. Each agent's state $x_i(t)$ is the karma score an agent accumulated as an engagement proxy. We perform four experiments, each tests one structural prior prediction against a null hypothesis $H_0$ based on current agentic AI systems.

\paragraph{Dataset Statistics}
\autoref{tab:dataset_summary} shows the data statistics of our the subset of MoltBook interactions that we analyzed. 

For P1 and P2, we partitioned agents by network degree into hub agents (25th- percentile), periphery agents (75th-percentile) and mid agents (the rest of the agents). This yielded groups of 4,571, 8,707, and 4,970 agents respectively.

\begin{table}[ht]
\centering
\small
\begin{tabular}{lccc}
\toprule
\textbf{Dataset} & \textbf{Posts} & \textbf{Comments} & \textbf{Unique authors} \\
\midrule
Full dataset & 290,251 & 1,836,711 & 39,700 \\
Analysis subset & 289,838 & 1,835,175 & 39,656 \\
\bottomrule
\end{tabular}
\caption{Summary statistics of the MoltBook dataset and analysis subset.}
\label{tab:dataset_summary}
\end{table}

\label{app:detailed_results}
\begin{figure}[h]
    \centering
    \includegraphics[width=0.8\linewidth]{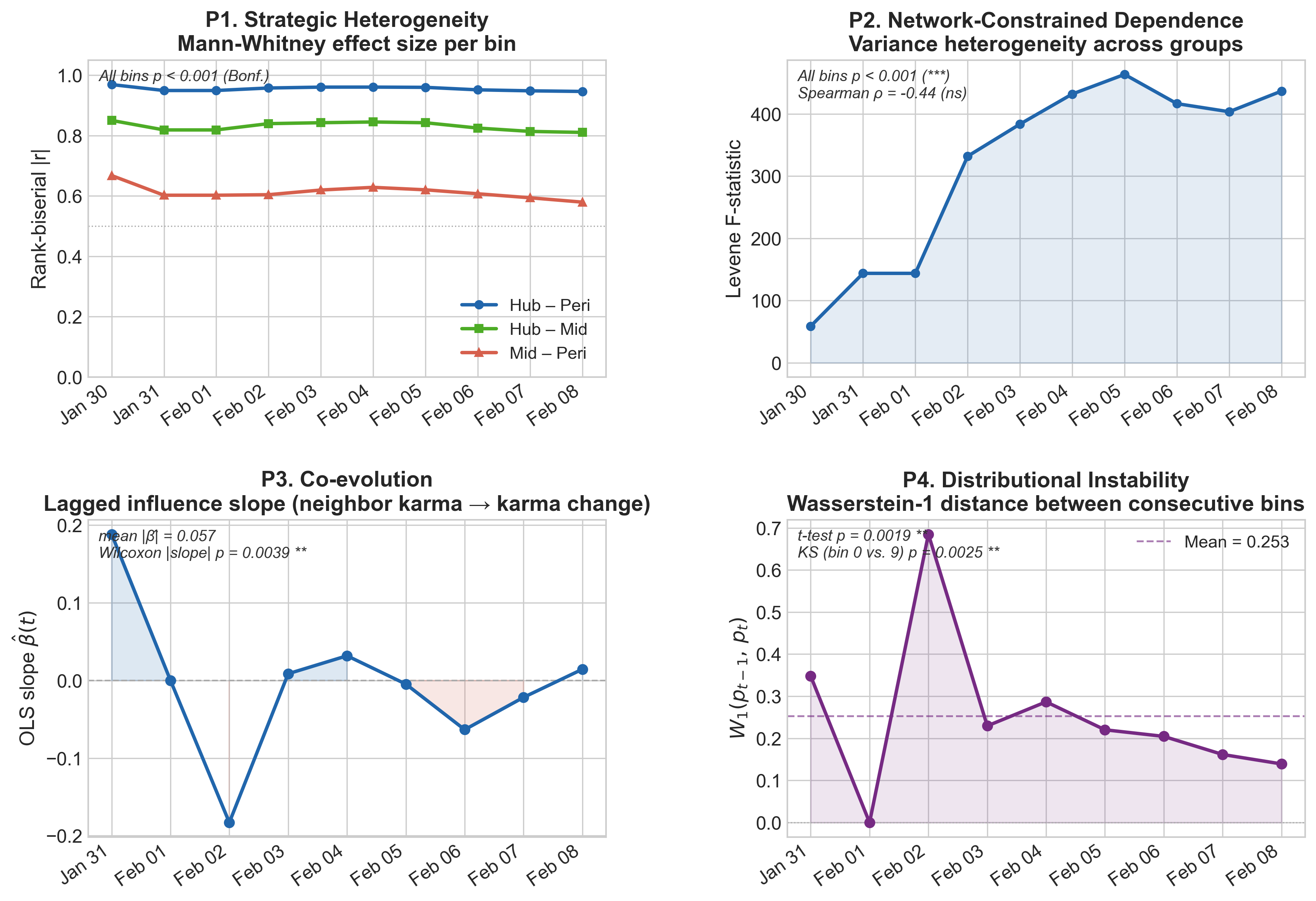}
    \caption{Statistical summary of MoltBook analysis}
    \label{fig:results}
\end{figure}

\paragraph{Network Visualization} Plotting all 18,248 nodes would render a network visualization undreadable, so we pruned the visualization to capture a representative view of the reply network's community structure. For the visualization, we plotted the reply network of the top $n=300$ most socially active agents, and clustered the network using a Louvain clustering algorithm. The nodes are colored by the log-transformed karma values. 

\subsection{Expt P1.} We plotted three time series of the mean log-transformed karma values of each group of partitioned agents. Specifically, we plotted: 
$$\text{P1}_g(t) = \frac{1}{|g|} \sum_{i \in g} |x_i(t)|, \quad g \in \{\text{hub, mid, peri}\}$$

We observe divergence in the groups, indicating that the network position determines engagement trajectories. 

\autoref{tab:kw_mwu_results} shows the Kruskal-Wallis tests of whether the mean of log-transformed karma differs significantly across the three agent partitions (Hub, Periphery, Mid) agents at each daily bin. Then, the Mann-Whitney U test performs pairwise comparison of Hub vs. Periphery agents. The Kruskal-Wallis tests confirm significant group differnces at every bin. Hub agents rank significantly higher in mean log-transform scores than Periphery agents at all time steps. 

\begin{table*}[ht]
\centering
\small
\resizebox{\textwidth}{!}{
\begin{tabular}{l l c c c c c c c}
\toprule
\textbf{Bin} & \textbf{Date} & \textbf{KW $H$} & \textbf{KW $p$} & \textbf{Sig} & \textbf{MWU $U$ (Hub-Peri)} & \textbf{MWU $p$} & \textbf{Bonf. sig} & \textbf{Effect $r$} \\
\midrule
0 & 2026-01-30 & 1090.69 & $1.44\times10^{-237}$ & *** & 183917 & $8.91\times10^{-126}$ & Yes & $-0.969$ \\
1 & 2026-01-31 & 2287.83 & $0.00$ & *** & 926534 & $5.74\times10^{-281}$ & Yes & $-0.949$ \\
2 & 2026-02-01 & 2287.83 & $0.00$ & *** & 926534 & $5.74\times10^{-281}$ & Yes & $-0.949$ \\
3 & 2026-02-02 & 5043.85 & $0.00$ & *** & 4493629 & $0.00$ & Yes & $-0.957$ \\
4 & 2026-02-03 & 5987.06 & $0.00$ & *** & 6149978 & $0.00$ & Yes & $-0.960$ \\
5 & 2026-02-04 & 7091.62 & $0.00$ & *** & 8334921 & $0.00$ & Yes & $-0.960$ \\
6 & 2026-02-05 & 7850.77 & $0.00$ & *** & 10541788 & $0.00$ & Yes & $-0.959$ \\
7 & 2026-02-06 & 8238.96 & $0.00$ & *** & 12527489 & $0.00$ & Yes & $-0.951$ \\
8 & 2026-02-07 & 8560.15 & $0.00$ & *** & 14356040 & $0.00$ & Yes & $-0.948$ \\
9 & 2026-02-08 & 9039.28 & $0.00$ & *** & 17548110 & $0.00$ & Yes & $-0.946$ \\
\bottomrule
\end{tabular}
}
\caption{Kruskal–Wallis (KW) and Mann–Whitney U (MWU) test results across bins. Bonferroni-corrected significance indicated. Significance levels: *** $p<0.001$.}
\label{tab:kw_mwu_results}
\end{table*}
\subsection{Expt P2.} We plotted three time series of the log-transformed karma variance of each group of partitioned agents. Specifically, we plotted: $$\text{P2}_g(t) = \mathrm{Var}_{i \in g}[x_i(t)]$$

The non-equal variance for each agent partition confirms that degree position changes exposure to engagement fluctuations. There is a higher variance for hub agents, followed by mid agents followed by periphery agents. 

We performed tests to determine whether the log-transformed variance of karma differs across each of the partitioned agents at each bin. \autoref{tab:p2_tests} is a Levene test with 3-way, and the center is the median, and the Fligner-Killeen test of Hub vs. Periphery. Both tests confirm that the log-transformed karma variance differs significantly across degree groups at every bin.

\begin{table}[ht]
\centering
\small
\begin{tabular}{lccc}
\toprule
\textbf{Test} & \textbf{Statistic} & \textbf{p-value} & \textbf{Sig} \\
\midrule
Wilcoxon signed-rank on $|\text{slope}|$ (one-tailed greater) 
& $W = 36.0$ & 0.0039 & ** \\

One-sample $t$-test on $|\text{slope}|$ vs.\ 0 (one-tailed) 
& $t = 2.289$ & 0.0257 & * \\
\bottomrule
\end{tabular}
\caption{Statistical tests on absolute slope values. Mean $|\hat{\beta}| = 0.0573$. Significance levels: * $p<0.05$, ** $p<0.01$.}
\label{tab:p2_tests}
\end{table}

\subsection{Expt P3.} For each step $t \in \{1, \ldots, n\}$, we performed the following steps:

1. Karma change: $\Delta x_i(t) = x_i(t) - x_i(t-1)$

2. Mean neighbor karma at $t-1$: $\bar{x}_{N(i)}(t-1) = \frac{1}{|N(i)|} \sum_{j \in N(i)} x_j(t-1)$. Neighbors are derived from the one-hop ego-network neighbors from the reply graph. Isolated agents are excluded. 

3. OLS regression with the equation $\Delta x_i(t) \sim \hat{\beta}(t) \cdot \bar{x}_{N(i)}(t-1) + \varepsilon$
$\hat{\beta}(t)$ measures how strongly the $t-1$ neighbor karma predicts $t$ change. 

Non-zero, time varying slopes ($\hat{\beta}(t)$) indicate that the reply graph transmit engagement signals across agents, establishing the graph-to-agent state portion of co-evolution. Since the graph itself is, by construction, endogenously formed by reply-driven interactions (agent-state-to-graph), these two portions together close the co-evolutionary loop: agent states shape the network, the network then shapes future agent states.
This is consistent with studies of network-mediated influence changes of co-evolution \cite{aral2009distinguishing}.

\autoref{tab:abs_slope_tests} shows the results where we test whether the OLS slopes $\hat\beta(t)$ have a non-zero magnitude. This slope measures the difference between the neighbor's karma at $t-1$ and the agent's karma change at $t$. Having a non-zero magnitude indicates that the reply network $G$ transmits engagement signals regardless of direction. Since the slopes oscillate in sign, the test is performed on $|\hat\beta(t)|$ rather than the signed $\hat\beta(t)$.

\begin{table}[ht]
\centering
\small
\begin{tabular}{lccc}
\toprule
\textbf{Test} & \textbf{Statistic} & \textbf{p-value} & \textbf{Sig} \\
\midrule
Wilcoxon signed-rank on $|\text{slope}|$ (one-tailed greater) 
& $W = 36.0$ & 0.0039 & ** \\

One-sample $t$-test on $|\text{slope}|$ vs.\ 0 (one-tailed) 
& $t = 2.289$ & 0.0257 & * \\
\bottomrule
\end{tabular}
\caption{Statistical tests on absolute slope values. Mean $|\hat{\beta}| = 0.0573$. Significance levels: * $p<0.05$, ** $p<0.01$.}
\label{tab:abs_slope_tests}
\end{table}

The observation of non-zero and time-varying slopes indicate that the reply network structure is predictive of individual karma change, consistent with network-mediated influence changes of co-evolution \cite{aral2009distinguishing}.

For completeness, \autoref{tab:p3_direction_test} tests for directional bias and trend of the $\hat\beta(t)$ coefficient. The signed slopes alternate in direction (fraction positive = 0.44), indicating bidirectional coupling, that is, there exist both positive and negative influence days, rather than a unidirectional trend, consistent with the mutual co-evolution claim.

\begin{table}[ht]
\centering
\small
\begin{tabular}{lcccc}
\toprule
\textbf{Test} & \textbf{Statistic} & \textbf{p-value} & \textbf{Sig} & \textbf{Notes} \\
\midrule
Wilcoxon signed-rank on $\hat{\beta}(t)$ vs.\ 0 (two-tailed) 
& $W = 18.0$ & 1.0000 &  & no net directional bias \\

One-sample $t$-test on $\hat{\beta}(t)$ vs.\ 0 
& $t = -0.103$ & 0.9203 & & 95\% CI [$-0.0776$, $0.0709$] \\

Spearman trend ($\hat{\beta}$ vs.\ bin) 
& $\rho = -0.233$ & 0.5457 &  & --- \\
\bottomrule
\end{tabular}
\caption{Statistical tests on slope estimates $\hat{\beta}(t)$. Summary statistics: mean $=-0.0033$, median $=0.0000$, fraction positive $=0.44$. }
\label{tab:p3_direction_test}
\end{table}

\subsection{Expt P4.} We computed the Wasserstein-1 distance between the distributions of log-transformed agent karma across time. We used the scipy.stats.wasserstein\_distance implementation. 
Specifically, we calculated: $$W_1(t) = W_1\!\bigl(p_{t-1},\, p_t\bigr)$$

The results show non-zero and time-varying $W_1$ which indicates endogenous distributional shift driven by interaction dynamics rather than a fixed external data distribution.

\autoref{tab:p4_stats_results} tests whether the 9 Wasserstein-1 distances (i.e., one distance for each time step) are all significantly greater than zero, which means whether the overall karma distribution shifted across the window. All tests confirm that consecutive daily karma distributions are non-stationary (mean $W_1=0.2528$). The KS test confirms that the overall distribution has shifted significantly from the first to the last observation day ($D=0.0456,p=2.5E-3$). These results are consistend with endogeneous distributional drift driven by agent interactions.

\begin{table}[ht]
\centering
\small
\begin{tabular}{lccc}
\toprule
\textbf{Test} & \textbf{Statistic} & \textbf{p-value} & \textbf{Sig} \\
\midrule
Wilcoxon signed-rank vs.\ 0 (one-tailed greater) & $W = 36.0$ & 0.0039 & **\\
One-sample $t$-test vs.\ 0 (one-tailed) & $t = 4.011$ & 0.0019 & ** \\
KS test: bin 0 vs.\ bin 9 & $D = 0.0456$ & $2.50\times10^{-3}$ & ** \\
Spearman trend ($W_1$ vs.\ bin) & $\rho = -0.483$ & 0.1875 &  \\
\bottomrule
\end{tabular}
\caption{Statistical test results for P4 Wasserstein-1 distance. Significance levels: ** $p<0.01$.}
\label{tab:p4_stats_results}
\end{table}

\section{Distinction with Adjacent Framings}
\label{sec:adjacent_framings}

\paragraph{Multi-agent Reinforcement Learning (MARL)} The MARL research program studies learned agents interacting under a shared reward structure, typically in simulated environments with well-defined payoffs \cite{wen2022multi}. MASS encompasses both real and simulated ecosystems where the rewards can differ across agents and can also be hidden (i.e., covert agents as often seen in state-sponsored agents on social media \cite{jacobs2023tracking}). 

\paragraph{LLM-based Agentic Frameworks} LLM-based agentic frameworks like AutoGen and MetaGPT \cite{hong2023metagpt,wu2024autogen} study interacting language agents collaborate on tasks like software engineering workflows and multi-turn problem solving. However, the agents in these systems collaborate under predefined roles and interaction protocols. Rather than focus on such micro-level agent-agent coordination, MASS emphasize the macro-level, where heterogeneous agents interact in an evolving network and result in emergence of population-level phenomena like information diffusion, influence and collective coordination \cite{cao2016evolutionary,ferrara2015quantifying,ng2023combined}. 

\paragraph{Theory-of-Mind (ToM)} The Theory-of-Mind to the capacity of an agent to model the beliefs, intentions and knowledge states of other agents \cite{sap2022neural}. The ToM research program asks whether agent $i$ can represent agent $j$'s belief state.
Social theory further understands ToM as socially situated \cite{smith2004socially}: an agent's model of others is filtered through $N(i)$ and shaped by the structural context of $G$. ToM is a component of $(f,g)$, and MASS is the system within which these components operate at scale. 

\paragraph{Social simulation} Generative agent simulation work can model populations with rich agent-agent interaction dynamics \cite{park2023generative,ng2025aurasight,anthis2025position}. As such, one might argue that MASS are already well captured within existing simulation frameworks, where agent behavior and interaction can be studied at scale.

We view social simulation and MASS as complementary rather than equivalent. Social simulation asks: can we model an agent's behavior realistically within a controlled environment? MASS asks: what happens when agents are deployed in open, real-world environments where they coexist and interact with humans and other agents? While simulation provides a valuable tool for understanding agent behavior, MASS focuses on the dynamics that arise when the model and the world are no longer disparate.

\paragraph{Social cybersecurity} The social cybersecurity research program has, among others, studied how automated bots accounts shape online discourse \cite{carley2020social}. MASS extends this lineage towards increasingly agentic bot-human systems. The detection, coordination and influence problems that the field has been studying do not disappear, but the study of heuristic-based and scripted bots would give way to understanding populations consisting of adaptive, AI-powered social actors.

\section{Design Principles \& Evaluation Benchmarks for MASS}
\label{app:principles_benchmarks}

\subsection{Design Principles for MASS}
Here are some design questions to guide the development of AI systems operating in a multi-agent and interactive environment. These principles are aligned along the four structural properties.

\paragraph{P1. Distributional Instability} This property states that the data distribution $p(x)$ is not fixed, and is generated by the system itself. Design questions to consider are:
\begin{itemize}
    \item What is the expected distribution or distribution shift of information $p(x)$ at deployment time?
    \item How do the agents handle feedback loops where their outputs influence their future inputs? The design should consider if the agent should be robust to such feedback loops or should adapt accordingly.
    \item Does the system maintain stability under evolving outputs? If not, does the system have a mechanism to detect when $p(x)$ has drifted beyond acceptable bounds? The design should have mechanisms like memory decay or diversity constraints to prevent runaway amplification. 
    \item Is your evaluation dataset drawn from the same distribution the system encounters at $t>0$? The design of the evaluation dataset must be similar to the distribution as the original system, or an analogous distribution.
\end{itemize}

\paragraph{P2. Co-evolution} This property states that the agents and environment mutually shape each other over time. Design questions to consider are: 
\begin{itemize}
    \item Do agent behaviors generate content, signals or actions that become inputs to other agents or to future versions of itself? If so, what is the feedback loop, and what is the timescale that it operates on?
    \item How do agent updates account for multi-round interaction effects? That is, how does the system adapt in response to the agents behavior? The system can react dynamically, or only under specific conditons.
    \item What happens when perturbations to the system is introduced? These perturbations can be new agents or policy changes.
    \item Is there a mechanism to detect and respond to adversarial adaptation (e.g. agents that exploit the original policy) or cascading feedback loops (e.g. echo-chambers, infinite cascades)? 
\end{itemize}

\paragraph{P3. Strategic Heterogeneity} This property states that agents differ in roles, incentive and influence.
\begin{itemize}
    \item What types of agents exist? The design must consider the different agents define their behavioral mechanics $\theta_i$. The design must also characterize the functional roles that these archetype play in $G$, i.e. are they a bridging node, a broadcast node, or a peripheral node? 
    \item How is influence distributed across agents? Influence can be distributed equally, or centralized at the hubs.
    \item Does the system behave appropriately when interacting with each archetype, or has it only been tested against a homogeneous agent population? 
    \item Can a particular archetype exploit the mechanics of $(f,g,G)$ that produces harmful population-level outcomes? An amplifier bot that extensively uses star broadcast structures can flood the network $G$ with undesired narratives, therefore only allowing its narrative to pass through $f$.
\end{itemize}

\paragraph{P4. Network-Constrained Dependence} This property states that interaction structures determine information and influence flow.
\begin{itemize}
    \item What is the interaction graph $G$? The interaction network can be defined through an artifact of the agent's behavior (i.e., replies), as is common with social media and communication studies. It can also be a fixed network that is determined apriori at the system's initialization.
    \item What does each agent observe? Who can observe whom, and who can influence whom? The design should consider the local vs global context, and what $N(i)$ looks like for the agents.
    \item How does the network topology (i.e., hubs, cluster, bridges, isolates) affect outcomes?
    \item What constraints exist on agent communication and visibility of information?
    \item Are there structural positions in $G$ that could give an agent disproportionate influence? The design must consider such types of positions and the agents that are likely to occupy these positions, and how the system is affected by such positional occupancy.
\end{itemize}

\subsection{Evaluation Benchmarks}
We outline example benchmark configurations that can operationalize evaluation under Multi-Agent Social Systems, focusing on population-level dynamics rather than individual task performance. These benchmarks are not intended as exhaustive or definitive, but intended to provide a starting point for the community to deploy and evaluate MASS.

\paragraph{Stability Under Interaction} This benchmark tests the system's robustness under distributional instability. 
\begin{itemize}
    \item Setup: Initialize $N$ agents with the corresponding diverse initial states $x_i$. Place agents in the network $G$. Run interaction mechanics for $T$ rounds. 
    \item Measures: distributional drift over time, convergence or divergence of outcomes and agent states $x_i$
\end{itemize}

\paragraph{Sensitivity to Perturbation} This benchmark tests co-evolution mechanics and system's robustness or fragility.
\begin{itemize}
    \item Setup: Initialize $N$ agents with the corresponding diverse initial states $x_i$. Place agents in the network $G$. Inject a small number of agents with biased behavior at time $t=k$. Run interaction mechanics for $T$ rounds.
    \item Measures: Shift in population distribution, information cascade size and spread (from $f$), time to recovery to original distribution, if any
\end{itemize}

\paragraph{Heterogeneity Stress Test} This benchmark tests the effects of strategic heterogeneity of agent archetypes.
\begin{itemize}
    \item Setup: Initialize $N$ agents in a network $G$ and run interaction mechanics for $T$ rounds. Vary the set of $N$ agents types: homogeneous vs heterogeneous populations, role-based agents (e.g. amplifiers, coordinators)
    \item Measures: outcome sensitivity to agent composition, role-dependent effects, patterns of information and influence over agent composition
\end{itemize}

\paragraph{Network Topology Variation} This benchmark tests the effects of network topology on the system.
\begin{itemize}
    \item Setup: Initialize $N$ agents in a network $G$ and run interaction mechanics for $T$ rounds. Run the same system under different network topologies of $G$, like erdos-renyi, small-world or scale-free graphs.
    \item Measures: cascade structure, diffusion speed, local vs global agreement, outcome variability with network topology
\end{itemize}


\newpage

\end{document}